\documentclass[journal]{IEEEtran}
\usepackage{amssymb}
\usepackage{bm}
\usepackage{subfigure}
\usepackage{algorithm}
\usepackage{algpseudocode}
\usepackage{multirow}
\usepackage{url}

\usepackage{stfloats}
\usepackage{xcolor}
\usepackage{ulem}
\usepackage{lettrine}
\usepackage{mathtools}
\usepackage{subfigure}
\usepackage{comment}
\usepackage{fancyhdr}
\usepackage{hyperref}

\fancypagestyle{firststyle}
{
   \fancyhf{}
   
   \fancyhead[C]{\scriptsize{This is the post-print of the following article: F. Rinaldi, S. Pizzi, A. Orsino, A. Iera, A. Molinaro and G. Araniti, ``A novel approach for MBSFN Area Formation aided by D2D Communications for eMBB Service Delivery in 5G NR Systems," in IEEE Transactions on Vehicular Technology, doi: 10.1109/TVT.2019.2958424. \\Article has been published in final form at: \url{https://ieeexplore.ieee.org/document/8928548}. \\Copyright~\copyright~2019 IEEE.}}
}
\ifCLASSINFOpdf
  \usepackage[pdftex]{graphicx}
\else
\fi
\begin{document}
%
\title{A novel approach for MBSFN Area Formation aided by D2D Communications for eMBB Service Delivery in 5G NR Systems}
%
%
%


\author{Federica~Rinaldi, \IEEEmembership{Student Member,~IEEE,}
        Sara~Pizzi, \IEEEmembership{Member,~IEEE,}
       Antonino~Orsino, \IEEEmembership{Member,~IEEE,}
        Antonio~Iera, \IEEEmembership{Senior Member,~IEEE,}
        Antonella~Molinaro, \IEEEmembership{Member,~IEEE,}
        and~Giuseppe~Araniti, \IEEEmembership{Senior Member,~IEEE}
\thanks{
The publication has been prepared with the support of the ``RUDN University Program 5-100''.

F. Rinaldi and S. Pizzi are with the DIIES Department, University Mediterranea of Reggio Calabria, Via Graziella Loc.
Feo di Vito, 89100 Reggio Calabria, Italy (e-mail: federica.rinaldi@unirc.it, sara.pizzi@unirc.it). 

A. Orsino is with Ericsson Research, Hirsalantie 11, 02420 Jorvas, Finland (e-mail: antonino.orsino@ericsson.com).

A. Iera is with the DIMES Department, University of Calabria, via Pietro Bucci, 87036 Arcavacata di Rende (Cosenza), Italy (e-mail: antonio.iera@dimes.unical.it).

A. Molinaro is with the DIIES Department, University Mediterranea of Reggio Calabria, Via Graziella Loc.
Feo di Vito, 89100 Reggio Calabria, Italy, and also with  Laboratoire des Signaux et Syst\`emes, CentraleSup\'elec--Universit\'e Paris-Saclay, France (e-mail: antonella.molinaro@unirc.it). 

G. Araniti is with the DIIES Department, University Mediterranea of Reggio Calabria, Via Graziella Loc.
Feo di Vito, 89100 Reggio Calabria, Italy, and also with Peoples’ Friendship University of Russia (RUDN University),
6 Miklukho-Maklaya St, Moscow, 117198, Russia (e-mail: araniti@unirc.it). }}
\maketitle

\begin{abstract}
Forthcoming 5G New Radio (NR) systems will be asked to handle a huge number of devices accessing or delivering ``resource-hungry'' and high-quality services. 
In view of this, the new 5G Radio Access Technology (RAT) aims to support, in next releases, Multimedia Broadcast/Multicast Service Single Frequency Network (MBSFN) to enable the simultaneous delivery of the same content to a set of users covered by different cells. 
According to MBSFN, all cells belonging to the same MBSFN Area are synchronized in time and the MBSFN transmission occurs over the same radio resources. In such a way, the same content flow is delivered by several cells to all the receivers in the MBSFN Area. 
A further means to enhance the network coverage and provide high data rate and low latency in future 5G-enabled MBSFN networks is Device-to-Device (D2D) connectivity.
Along these lines, in this paper we propose a D2D-aided MBSFN Area Formation (D2D-MAF) algorithm to dynamically create MBSFN Areas with the aim to improve the system aggregate data rate while satisfying all user requests. The proposed D2D-MAF foresees that users could receive the service through either MBSFN, or D2D, or unicast transmissions. Performance evaluation results, carried out under a wide range of conditions, testify to the high effectiveness of the proposed algorithm.
\end{abstract}

\begin{IEEEkeywords}
5G, New Radio, eMBB, MBSFN, MBSFN Area Formation, D2D.
\end{IEEEkeywords}

\thispagestyle{firststyle}

%
\IEEEpeerreviewmaketitle

\section{Introduction} 
\label{introduction}

\IEEEPARstart{N}{OWADAYS}, the demand for resource-hungry services, such as mobile TV, video streaming and other multicast applications, is constantly growing and it is also encouraged by an ever-increasing number of devices that require connection to the Internet. Cisco VNI forecast \cite{48} highlights that mobile video traffic will increase 9-fold between 2016 and 2021, by accounting for 60\% of the total mobile data traffic in 2016, and achieving 78\% of the world mobile data traffic by 2021.
Moreover, Ericsson Mobility Report \cite{EricssonMobilityReport} predicts that the number of smartphone subscriptions will increase by 45\% by generating 95\% of total mobile data traffic at the end of 2024.
Projecting towards the near future, the next-to-come fifth-generation wireless system (i.e., 5G) promises to meet the needs of such an exigent world market. 

In order to manage the possibility of a huge number of users demanding for the same service, the 5G system, also known as 5G ``New Radio (NR)'' because of its new air interface \cite{3gpp_rel14}, in its future releases should introduce group-oriented communications, already supported by LTE thanks to evolved-Multimedia Broadcast/Multicast Service (eMBMS) \cite{eMBMS} (i.e.,the evolution of MBMS \cite{2}). eMBMS exploits Point-to-Multipoint (PtM) transmissions to deliver data to a group of users interested in receiving the same service. In this regard, recent releases of LTE introduced the possibility to perform a dynamic resource allocation between unicast and multicast traffic. Since Release 14 \cite{3gpp_rel14}, the constraint to allocate at most 60\% of the available resource has been removed and this allows now to assign up to 100\% of the available radio resources to multicast/broadcast.

Among the most interesting eMBMS features there is MBMS over Single Frequency Network (MBSFN) \cite{7}, which could be an appealing solution also for 5G NR technology to cover large geographical areas wherein nearby NR base stations (gNodeBs or gNBs) could be synchronized in time to deliver the same content over the same radio resources. An MBSFN transmission is enabled by involving gNBs belonging to the same MBSFN Area \cite{11}, wherein the multiple replicas of the signal from more cells, seen as multi-path components of the same signal from a single cell, may be constructively combined at the receiver with a consequent improvement in the perceived Quality of Service (QoS).

Along the evolutionary path from High-Speed Packet Access (HSPA) to Long Term Evolution-Advanced (LTE-A), the process of MBSFN Area formation has always followed a static approach \cite{46}. This implies that all adjacent cells interested in a given service are grouped into a single MBSFN Area and the content is broadcasted with the most robust modulation supported by users via a single MBSFN transmission.
Differently, the next-generation Radio Access Technologies (RAT) \cite{3gpp_rel14} foresees a dynamic adjustment of the MBSFN Area according to, for example, required services, user distribution, or a given cost function. Furthermore, as for LTE, the new RAT supports the coexistence among unicast and Multicast/Broadcast transmissions towards users.

As a consequence of the LTE evolution, additional enhancements have boosted the system capacity and user data rates. Indeed, Device-to-Device (D2D) communications were introduced in the LTE Release 12. The term D2D refers to the direct connection through sidelinks between devices in mutual proximity, without involving the eNB (the LTE base station), finalized to improve cellular coverage, increase throughput, and reduce energy consumption and delay. Over and above these important features, the more relevant novelty is that D2D revolutionizes the traditional transmission mode (from base stations to devices) by establishing an innovative and unconventional communication \cite{Mancuso_d2d} modality.
For the aforementioned achievable benefits, D2D communications will play a key role also in the forthcoming 5G NR system. 

On top of this, the 3rd Generation Partnership Project (3GPP) standardization efforts towards 5G have involved also the existing LTE radio access technology. As a result, LTE Release 15 represents a fundamental piece in the 5G puzzle and is rightly considered as the NR predecessor. The enhancements done in such technology involve both the core and the radio access network. One of the most relevant is the possibility for the eNBs to be connected to the new 5G core network (5GC); thus allowing the LTE eNBs to inherit part of the new 5G NR functionalities. 


In this paper, we propose a D2D-aided MBSFN Area formation algorithm, hereinafter referred to as D2D-MAF, that determines the best configuration of MBSFN Areas in terms of system aggregate data rate {(ADR)}. The proposed approach aims to offer a higher quality of service to users within the MBSFN Area w.r.t. the static approach and, at the same time, to improve the quality of the service delivered to those users experiencing poor channel conditions. In particular, D2D-MAF first groups adjacent cells requesting the same MBMS service under an MBSFN Area wherein the MBSFN transmission occurs with the most robust Modulation and Coding Scheme (MCS) among those supported by the users under coverage. Afterward, the MCS level is iteratively increased and, in case not all devices can decode the broadcasted content via the multicast transmission, D2D-MAF exploits both unicast link and the D2D connectivity to satisfy the requests of users not served by the MBSFN Area. In detail, if a device can receive the service through a direct connection with a forwarding device belonging to the MBSFN Area, then D2D connections are established; otherwise, that content is provided through a unicast transmission. Hence, in the same cell the MBMS service can be delivered through either MBSFN,  D2D, or unicast transmission.
To assess the effectiveness of the proposed D2D-MAF, simulation campaigns have been carried out and results compared with those obtained by the Single Content Fusion (SCF) scheme \cite{8} \cite{SCF_2017}, which we consider as a benchmark. The system performance analysis has been carried out under different TDD frame configurations for both D2D-MAF and SCF.

It is worth emphasizing once again that MBSFN and D2D communications are currently not supported in 5G NR (i.e., work on this technologies will start in Release 16 and 17) and, thus, the intent of this paper is also to initiate a research activity that can be a starting point for future academic and industrial investigations on features, mechanisms, and solutions needed to integrate multicast and D2D transmissions into 5G NR.

The remainder of the paper is organized as follows.
Section \ref{related work} briefly overviews MBSFN Area formation solutions. Section \ref{system model} describes the reference system model, while our proposed D2D-MAF algorithm is presented in section \ref{D2D-MAF}. The simulative model and simulation results are shown in section \ref{performance evaluation}. Finally, conclusions are drawn in Section \ref{conclusion}.

\section{Related works}
\label{related work}

Multicast and broadcast transmissions will play a key role in 5G wireless systems \cite{IEEENetwork}. Indeed, next-generation wireless systems will support a wide variety of enhanced traditional applications, such as breaking news, alerts in emergency scenarios, live streaming of popular music or sport events, and software updates. Moreover, new multicast/broadcast services (i.e., eMBMS \cite{eMBMS}, location/position-based and critical communication services \cite{ref_standard_pos}, and Vehicle-to-Everything (V2X) services \cite{ref_standard_Mol1}, \cite{ref_standard_Mol2}) are definitely gaining ground. 

To increase 5G system flexibility in guaranteeing all mentioned services, the International Telecommunication Union (ITU) \cite{ITU} defines three usage scenarios, featured by different requirements.
\textit{Enhanced mobile broadband (eMBB)} requires high data rates across a wide coverage area.
\textit{Massive Machine Type Communications (mMTC)} need low energy consumption to support connections among a huge number of low-cost, low-power and long-life devices.
Finally, \textit{Ultra-Reliable and Low Latency Communications (URLLC)} allow bidirectional communications among devices requiring low latency and high network reliability.
In this paper, we refer to an eMBB scenario. 


In the case of multicast transmissions, the Adaptive Modulation and Coding (AMC) RRM technique determines the transmission parameters on a per-group basis and chooses the proper MCS according to the channel qualities perceived by all users.
In the literature, RRM schemes can be classified into single-rate and multi-rate approaches.
The former includes the Conventional Multicast Scheme (CMS) \cite{CMS} and the Opportunistic Multicast Scheme (OMS) \cite{OMS}.
The principle of CMS is to broadcast the content to all multicast users with the lowest MCS, thus suffering from poor spectral efficiency.
Differently, under OMS the content is delivered only to the best set of users every Transmission Time Interval (TTI). Although achieving long-term fairness, OMS suffers from short-term unfairness. 
Finally, the Multicast Subgrouping (MS) \cite{25} is a multi-rate approach that splits users into subgroups, each characterized by a different MCS. MS offers a good trade-off between fairness and throughput.
 
To the best of our knowledge, still only a few works in the literature dealt with the problem of MBSFN Area formation and its related issues (i.e., how to group cells and which contents to broadcast). 
The 3GPP standard foresees two techniques for MBSFN Area formation. One is represented by the static approach, also known as Legacy \cite{46}, which statically includes all cells synchronized for performing the MBSFN transmission in a single MBSFN Area. 
Another approach is Single Cell Point-to-Multipoint (SC-PtM) \cite{47}, which supports the multicast transmission over single cell and the MBSFN Area is adjusted cell by cell. However, SC-PtM does not benefit from the advantages of Single Frequency Networks, which is from the possibility to receive multiple replicas of the same content from different base stations within a certain time interval (i.e., cyclic prefix).

The Single Content Fusion (SCF) scheme \cite{8} \cite{SCF_2017} is the first work providing a dynamic method for MBSFN Areas formation and content items selection, based on interest similarity. SCF first creates single-content MBSFN areas, including cells with similar content interests, and then it merges the created MBSFN Areas that could overlap. SCF objective is to create multi-content MBSFN Areas wherein to broadcast the most demanded contents that maximize the overall throughput by exploiting both MBSFN and unicast transmissions. More in detail, SCF increases the minimum MCS level for the MBSFN transmission and serves the users with poor channel conditions via unicast links. 
Another dynamic solution for the problem of MBSFN Area formation is proposed in \cite{9}, where the display capabilities of devices are considered in order to maximize the quality of experience of users in the MBSFN Area.
Further works, \cite{6} and \cite{7.1}, deal with MBSFN. In \cite{6}, the authors propose an algorithm that maximizes the coverage and achieved results show that the greater the number of areas the better the system performance. On the other hand, \cite{7.1} analyzes the performance when increasing the size of the MBSFN Area and varying the separation among eNBs. In particular, the higher the minimum separation among eNBs, the better is the achieved performance. Although \cite{6} and \cite{7.1} deal with MBSFN, they do not focus on the problem of MBSFN Area formation; rather, they address the issue of the best configuration of MBSFN Areas in terms of performance.

In this paper, we design a heuristic algorithm to solve the problem of MBSFN Area formation by \textit{(i)} searching for the best MBSFN configuration to increase the system aggregate data rate, \textit{(ii)} exploiting both unicast and D2D communications to deliver the eMBB content to cell-edge users, \textit{(iii)} evaluating the system performance on a per-frame basis, \textit{(iv)} performing a proper radio resource allocation to MBSFN, unicast, and D2D transmissions, and \textit{(v)} delivering the eMBB service to all the interested users.
With respect to the literature in the field, a major novelty introduced by our work is the exploitation of D2D communication to improve the performance of the MBSFN Area formation process. While in \cite{8} the authors plan to establish unicast connections to create multi-content MBSFN Areas wherein to broadcast the most demanded contents, according to our approach this is considered ``the last chance’’ to adopt just in case it is not possible to take advantage from the creation of D2D links.

\section{System Model}
\label{system model}

We consider a Multicast/Broadcast Single Frequency Network (MBSFN) in 5G NR systems, where all gNBs are time-synchronized and simultaneously deliver a service by exploiting the same frequency.
Although the emerging 5G system is characterized by a new OFDM-based air interface, known as New Radio \cite{nr}, its flexible design can guarantee an efficient coexistence with LTE when operating in the same licensed frequency band \cite{nrRAT}.

In the considered 5G NR system, downlink and uplink transmissions are organized in frames, each consisting of 10 subframes. Each subframe has a duration of 1 ms and is composed of a different number of slots depending on the numerology used \cite{nr}.
The available radio spectrum is managed in terms of Resource Blocks (RBs), each representing the smallest bandwidth needed to establish a transmission. An RB consists of 12 consecutive subcarriers, equally spaced according to the subcarrier spacing (SCS) determined by the numerology.

{We refer to an MBSFN scenario, where users are interested in the same service and cells broadcast the demanded content (i.e., a software update or a video stream).
In the official 3GPP study \cite{use_case_mbms}, several use cases for the eMBMS architecture are defined, which is live video from multiple camera angles into a stadium, nation-wide TV channels, Video on Demand (VoD) prepositioning, software update, and TV program guide update delivery. In this paper, we consider VoD scenarios in which multiple users require videos more or less simultaneously, such as, for example, videos of sport events in a stadium, guided tours inside a museum, or lectures in a classroom.}
Since VoD services require a wide channel bandwidth, we consider the numerology $\mu$=0, where the SCS is equal to 15 kHz and the TTI is set to 1 ms.

Let $\mathcal{C}$ be the set of cells (i.e, gNBs) deployed within a Synchronization Area, where a set of users $\mathcal{U}$ is interested in the same multimedia content.  A Synchronization Area may include one or more MBSFN Areas.  The set of all MBSFN Areas is denoted by $\mathcal{M}$. 
Each MBSFN Area $m \in\mathcal{M}$ consists of adjacent cells involved in an MBSFN transmission, where the same video content is delivered at the same time over the same set of radio resources.

We consider that the requested video content can be delivered to users by exploiting two transmission modes. The \textit{cellular mode} occurs when UEs receive the content directly from the gNB either in MBSFN Transmission or via unicast links, whereas the \textit{D2D mode} occurs when UEs receive the content from a forwarding UE through D2D communications without establishing any radio links with the gNB. 
Therefore, we define the set of users receiving the requested service through MBSFN transmission with $\mathcal{U}_b$, the set of users served via unicast with $\mathcal{U}_u$ and the set of users served in D2D with $\mathcal{D}$. Moreover, $\mathcal{R}$ refers to the set of relay users, i.e., forwarding UEs.
Each UE transmits its Channel State Information (CSI), including its Channel Quality Indicator (CQI), to the gNB, which performs the scheduling procedures by properly selecting the transmission parameters (i.e., MCS) and managing the pool of available radio resources at every TTI. We denote as $\mathcal{RB}_m$ the whole set of available RBs for the \textit{m}-th MBSFN Area. 
{As for the MBSFN and unicast transmissions, wherein the gNB is responsible for radio resource management performed on the basis of the previously received users’ CSI feedbacks, we assume that the gNB also configures transmission parameters of D2D connections (i.e., NR Sidelink Mode 1 \cite{v2x_synchronization}) before starting the MBSFN transmission from  the gNB to users included in the MBSFN Area. Hence, }
$\mathcal{RB}_b$, $\mathcal{RB}_u$ and $\mathcal{RB}_{D2D}$ indicate the sets of RBs dedicated to the MBSFN, unicast, and D2D transmissions, respectively.
We also assume that D2D communications occur in Single Frequency mode, i.e., all forwarding UEs in $\mathcal{R}$ {are synchronized in time and} cooperate to perform an MBSFN transmission towards UEs in $\mathcal{D}$ by simultaneously delivering, over the same bandwidth, the content previously received from the gNB. 
{
We highlight that, in our scenario, we consider ``network-assisted'' D2D connections in NR Sidelink Mode 1 \cite{v2x_synchronization}, where the gNB works as a central coordinator of the transmissions, in terms of both time synchronization and resource allocation.}

Moreover, D2D communications take place in uplink subframes in order to guarantee a more efficient radio resources reuse \cite{d2d-ul}. Differently, both MBSFN and unicast transmissions take place in downlink subframes, by sharing available resources. The main notations used in the proposed D2D-MAF algorithm are summarized in Table \ref{tab:note}.

\begin{footnotesize}
\begin{table}[ht]
\caption{Notation used in the proposed D2D-MAF algorithm}\label{tab:note}
\centering
{
\begin{tabular}{|l|l|}
\hline
$\mathcal{C}$ & set of cells within the Synchronization Area\\
\hline
$\mathcal{M}$ & set of all MBSFN Areas\\
\hline
$\mathcal{P}$ & set of cells being part of MBSFN Areas\\
\hline
$\mathcal{L}$ & set of CQI levels\\
\hline
$\mathcal{U}$ & set of users requiring the broadcasted service\\
\hline
$\mathcal{U}_j$ & set of users belonging to the \textit{j}-th cell\\
\hline
$\mathcal{U}_l$ & set of users with the \textit{l}-th CQI\\
\hline
$\mathcal{U}_u$ & set of users receiving the MBMS content via unicast link\\
\hline
$\mathcal{U}_b$ & set of users served through MBSFN transmission\\
\hline
$\mathcal{R}$ & set of relay devices\\
\hline
$\mathcal{D}$ & set of devices receiving the content via D2D link\\
\hline
$\mathcal{RB}$ & set of available resources\\
\hline
$\mathcal{RB}_{b}$ & RBs destined to the MBSFN transmission \\
\hline
$\mathcal{RB}_{u}$ & RBs dedicated to unicast \\
\hline
$\mathcal{RB}_{D2D}$ & RBs for establishing D2D communications\\
\hline
$\mathcal{ADR}$ & Aggregate Data Rate of the overall system\\
\hline
$\mathcal{ADR}_{B}$ & Aggregate Data Rate of users belonging to $\mathcal{U}_b$\\
\hline
$\mathcal{ADR}_{U}$ & Aggregate Data Rate of users belonging to $\mathcal{U}_u$\\
\hline
$\mathcal{ADR}_{D2D}$ & Aggregate Data Rate of users belonging to $\mathcal{U}_{D2D}$\\
\hline
\end{tabular}
}
\end{table}
\end{footnotesize}

The proposed D2D-MAF algorithm exploits unicast transmissions and D2D communications for dynamically adjusting the MBSFN Area coverage wherein to broadcast the eMBB content. In particular, the MBSFN Area delivers the requested service to users with good channel conditions. Among them, the D2D-MAF algorithm selects the users that will act as relay nodes to forward the content to other users via D2D transmissions. In order to establish a D2D communication, the main requirement is the mutual proximity among involved users. Finally, the remaining users, typically with worse channel conditions, receive the content via unicast links.

The proposed D2D-MAF algorithm must meet several constraints, briefly discussed below.

According to the 3GPP standard \cite{x}, a cell can belong to at most 8 MBSFN Areas and the number of MBSFN Areas within a synchronization area cannot exceed 256.

  {In the $m$-th MBSFN Area, all users interested in the broadcasted content shall be served through the MBSFN transmission or unicast links or D2D connections:
\begin{equation}\label{0}
  \mid\mathcal{U}_m\mid=\mid\mathcal{U}_{b,m}\mid+\mid\mathcal{U}_{u,m}\mid+\mid\mathcal{D}\mid, \quad \forall m \in \mathcal{M}
\end{equation}}

In the downlink subframe, the sum of RBs allocated to the MBSFN transmission and those assigned to the unicast users, interested in the broadcasted content, shall not exceed the number of available RBs in a given MBSFN Area:
  
\begin{equation}
(\mid \mathcal{RB}_{b}\mid + \mid\mathcal{RB}_{u}\mid) \le \mid\mathcal{RB}_{m}\textsuperscript{DL}\mid, \quad \forall m \in \mathcal{M}
\label{3}
\end{equation}

Finally, in the uplink subframe, the number of RBs allocated to D2D communications shall be at most equal to the number of available RBs in a given MBSFN Area:
 
\begin{equation}
\mid \mathcal{RB}_{D2D}\mid \le \mid\mathcal{RB}_{m}\textsuperscript{UL}\mid, \quad \forall m \in \mathcal{M}
\label{4}
\end{equation}

The purpose of the proposed D2D-MAF algorithm is to create MBSFN Areas that meet the above constraints and to efficiently assign radio resources in order to increase the overall ADR of all MBSFN Areas.

By referring to a synchronization area, consisting in the set $\mathcal{M}$ of MBSFN Areas, the ADR is given by:

\begin{equation}
\mathcal{ADR}=\sum\limits_{m\in\mathcal{M}}\bigl (\mathcal{ADR}_{B}+\mathcal{ADR}_{U}+\mathcal{ADR}_{D2D} \bigr)
\label{5}
\end{equation}

where:

\begin{itemize}
\item the aggregate data rate of users receiving the content through  the MBSFN transmission is represented as:
\begin{equation}    
\small{\mathcal{ADR}_{B}= \sum\limits_{b\in\mathcal{U}_{b{,m}}}Rate(b)\times \mid\mathcal{RB}_{b{,m}}\mid, \forall m \in \mathcal{M}},
\label{7}
\end{equation}
{where $Rate(b)$ is the data rate per RB related to the lowest (i.e., most robust) MCS level supported by all users in $\mathcal{U}_{b,m}$ receiving the content by the $m$-th MBSFN Area.}
\item sum of data rates achieved by all users served in unicast is equal to: \begin{equation}    
\small{\mathcal{ADR}_{U}= \sum\limits_{u\in\mathcal{U}_{u{,m}}}Rate(u)\times \mid\mathcal{RB}_{u{,m}}\mid, \forall m \in \mathcal{M}},
\label{8}
\end{equation}
{where $Rate(u)$ is the data rate per RB related to the MCS level supported by the $u$-th user in $\mathcal{U}_{u,m}$ receiving the content via unicast link in the $m$-th MBSFN Area.}
\item the aggregate data rate of users receiving the video content through D2D communications can be written as:
\begin{equation}    
\small{\mathcal{ADR}_{D2D}= \sum\limits_{d\in\mathcal{D}_{{m}}}Rate(d)\times \mid\mathcal{RB}_{D2D{,m}}\mid, \forall m \in \mathcal{M}},
\label{9}
\end{equation}
{where $Rate(d)$ is the data rate per RB related to the lowest (i.e., most robust) MCS level supported by all D2D users in $\mathcal{D}_{m}$ receiving the content through D2D links in the $m$-th MBSFN Area.}
\end{itemize}

Since the MBSFN Area formation is a NP-hard problem, D2D-MAF aims to solve, through a heuristic approach, the following problem:

\begin{equation}    
\underset{\mathcal{RB}} {\operatorname{arg\,max} \mathcal{ADR}}
\label{10}
\end{equation}
\\
\centerline {subject to  \eqref{0} - \eqref{5}}
\\
\\

{The NP-hardness is mainly due to the Hill Climbing phase, where the constraints on radio resources and on adjacency among cells are verified in order to first identify the set of MBSFN Areas and, then, compute the system ADR. In realistic contexts, this phase is a complex problem for the ADR maximization since, as shown in \cite{hill_climbing}, it requires very high computational costs. Therefore, we followed a heuristic approach to solve the problem with a lower complexity by increasing as much as possible the system ADR, but without finding the optimal value. }

Differently from LTE, which supports a fixed and rigid TDD frame configuration \cite{11}, in NR the TDD uplink-downlink frame configuration becomes dynamic thanks to the introduction of flexible subframes that can be selectively used either in the downlink or in the uplink. Hence, the flexible frame structure \cite{flexible_frame_structure} ensures the forward compatibility with LTE.
In this paper, we exploit Dynamic Spectrum Sharing (DSS) \cite{ericsson_spectrumsharing} to analyze the system performance when running NR by considering the TDD configurations shown in Table \ref{tab:frames}.

\begin{table}[ht]
\centering
\caption{TDD Frame Configurations (U: Uplink; D: Downlink).} 
\label{tab:frames}
\begin{tabular}{ccccccccccc}
\hline
\hline
\multirow{2}{*}{\bf Configuration} & \multicolumn{10}{c}{\textbf{Subframe Number}} \\
\cline{2-11}
 & \textbf{0} & \textbf{1} &\textbf{ 2} & \textbf{3} & \textbf{4} & \textbf{5} & \textbf{6} & \textbf{7} & \textbf{8} & \textbf{9} \\
\hline
\hline
0 & D & S & U & U & U & D & S & U & U & U\\
1 & D & S & U & U & D & D & S & U & U & D\\
2 & D & S & U & D & D & D & S & U & D & D\\
3 & D & S & U & U & U & D & D & D & D & D\\
4 & D & S & U & U & D & D & D & D & D & D\\
5 & D & S & U & D & D & D & D & D & D & D\\
6 & D & S & U & U & U & D & S & U & U & D\\
\hline
\hline
\normalsize
\end{tabular}
\end{table}

{Each TDD frame configuration consists of ten subframes that could be Uplink (U), Downlink (D), or Special (S). The number of U, D and S subframes varies from a TDD frame configuration to another. Moreover, the positions of U, D and S subframes within a frame also change for different TDD frame configurations.}
{Hence, the performance of the proposed D2D-MAF scheme is affected by the number and the positions of both U and D subframes within the frame. The reason for this is that D2D-MAF is designed so that both MBSFN and unicast transmissions occur in D subframes, while D2D communications occur in U subframes. It is worth noting that the data forwarding from a relay device to other devices through D2D transmission in U subframes happens only after the relay device has received the content from the gNB in D subframes.}

\section{D2D-MAF Algorithm}
\label{D2D-MAF}

The proposed D2D-MAF algorithm dynamically creates MBSFN Areas, whose transmission parameters (i.e., MCS) are set in such a way as to maximize the system aggregate data rate. 
To achieve this goal, the D2D-MAF algorithm exploits both unicast and D2D communications for data delivering to users not able to support the MCS selected for the MBSFN transmission.
Hence, the main idea is to raise the MCS of the MBSFN Area as much as possible; subsequently, the proposed algorithm searches for relay nodes among the $\mathcal{U}_b$ users. 
Once relays are found, users excluded from the MBSFN Area are classified into two categories. 
The first includes \textit{D2D users} receiving the content via D2D links from the selected relays, which are coordinated and time-synchronized to perform a single-frequency-based D2D communication {(i.e., D2D communication in single frequency mode) by re-using the same set of uplink radio resources when simultaneously forwarding the content to D2D users}. The second includes \textit{unicast users} receiving the service directly from the gNB over Point-to-Point links.

In order to guarantee efficient resource reuse, the D2D-MAF algorithm foresees that D2D communications occur in uplink subframes over all available radio resources. Unicast transmissions share with the MBSFN transmission the radio spectrum in downlink subframes, where the set of available radio resources are allocated in such a way as to avoid interference between MBSFN and unicast transmissions.

\begin{figure} [ht!]
\centering%
{\includegraphics[scale=0.4]{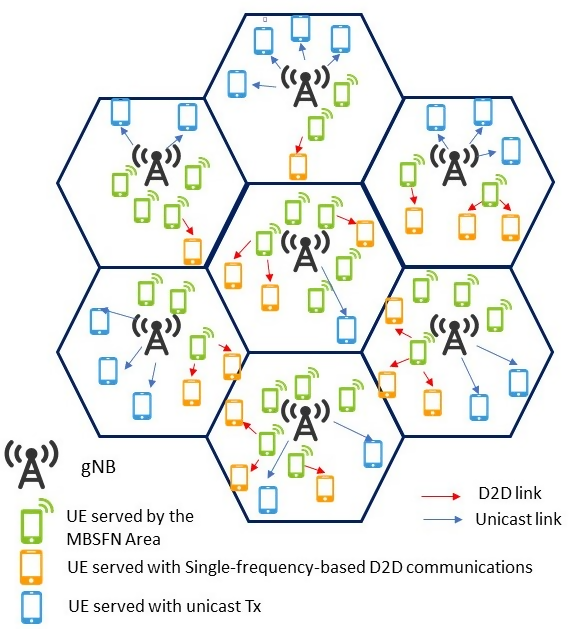}}
\caption{{Example of D2D-MAF scenario.}}
\label{fig:ibridoscenario}
\end{figure}

The objective of the proposed algorithm is to define the best configuration of MBSFN Areas and, at the same time, to select the set of multicast, unicast, and D2D users (i.e., $\mathcal{U}_{b}$, $\mathcal{U}_{u}$ and $\mathcal{U}_{D2D}$, respectively) that increase, as much as possible, the system aggregate data rate. Moreover, the D2D-MAF algorithm always meets the constraint to satisfy all user requests and, in the case of unicast transmissions, it performs radio resource allocation by following Round-Robin scheduling. 
In Figure \ref{fig:ibridoscenario}, we depict the operation of the proposed D2D-MAF algorithm. 

\subsection{Step-by-step implementation}

The pseudo-code that describes the operation of the proposed D2D-MAF algorithm is listed in Algorithm 1. 

\begin{algorithm}[ht]
\caption{D2D-MAF}\label{pseudocode1}
\begin{algorithmic}[1]

\begin{footnotesize}
\For {$\mathcal{U}_{{j}} \in \mathcal{U} $}
\If {$\mid \mathcal{U}_{{j}}\mid == 1$ }
\State {\textbf{add} $\mathcal{U}_{{j}}$ to $\mathcal{U}_{u}$; }
\ElsIf {$\mid \mathcal{U}_{{j}}\mid \ge 2$ }
\State {\textbf{add} $\mathcal{C}_{j}$ to $\mathcal{P}$; }
\State {\textbf{add} $\mathcal{U}_{{j}}$ to $\mathcal{U}_{b}$; }
\EndIf
\EndFor
\If {IsAdjacent($\mathcal{P}$)==TRUE}
\State {$\mathcal{M}=CreateMBSFNArea(\mathcal{P})$; }
\EndIf
\State {$\mathcal{ADR}=\mathcal{ADR}_{B}+\mathcal{ADR}_{U}$; }
\State {$\mathcal{L}=OrderUsrMCS(\mathcal{U}_b)$; }
\For {$l \in \mathcal{L}$}
\State {$\mathcal{U}_l=FindUsr(\mathcal{U}_b, l)$; }
\State {$\tilde{\mathcal{U}}_b=\mathcal{U}_{b} \setminus \mathcal{U}_{l}$; }
\For {$m_s \in \mathcal{M}$}
\For {$c_j \in \mathcal{C}$}
\If {$\mid \mathcal{U}_{l}\mid \ge 1$ }
\State {$\tilde{\mathcal{R}}=FindRelay(\tilde{\mathcal{U}}_b)$; }
\State {$\tilde{\mathcal{D}}=FindD2D(\mathcal{U}_{l}$); }
\State {$\tilde{\mathcal{U}}_u=\mathcal{U}_{l} \setminus \tilde{\mathcal{D}}$; }
\If {$\mid \tilde{\mathcal{R}}\mid \ge 1$ }
\State {\small{$\tilde{\mathcal{ADR}}(c_j)=\mathcal{ADR}_{B}(c_j)+\mathcal{ADR}_{D2D}(c_j)$; }}
\If { $\mid \tilde{\mathcal{U}}_u\mid > 0$ }
\State {\small{$\tilde{\mathcal{ADR}}(c_j)=\tilde{\mathcal{ADR}}(c_j)+\mathcal{ADR}_{U}(c_j)$; }}
\EndIf
\ElsIf {$\mid \tilde{\mathcal{R}}\mid == 0$ }
\State {\small{$\tilde{\mathcal{ADR}}(c_j)=\mathcal{ADR}_{B}(c_j)+\mathcal{ADR}_{U}(c_j)$; }}
\EndIf
\ElsIf {$\mid \mathcal{U}_l\mid == 0$ }
\State {\small{$\tilde{\mathcal{ADR}}(c_j)=\mathcal{ADR}_{B}(c_j)$; }}
\EndIf
\EndFor
\State{$\tilde{\mathcal{ADR}}=\sum\limits_{c_j\in {m}_{s}}\tilde{\mathcal{ADR}}(c_j)$; }
\EndFor
\State{$\tilde{\mathcal{ADR}}=\sum\limits_{m_s\in \mathcal{M}}\tilde{\mathcal{ADR}}(m_s)$; }
\If {$\tilde{\mathcal{ADR}} \ge \mathcal{ADR}$}
\State{$\mathcal{ADR}=\tilde{\mathcal{ADR}}$;}
\State {$\mathcal{U}_{b} =\tilde{\mathcal{U}}_b$; }
\State {$\mathcal{U}_{u} =\tilde{\mathcal{U}}_u$; }
\State {$\mathcal{R} =\tilde{\mathcal{R}}$; }
\State {$\mathcal{D} =\tilde{\mathcal{D}}$; }
\Else
\State{break;}
\EndIf
\EndFor
\end{footnotesize}
\State {\small{\textbf{return:} $\mathcal{ADR}, \mathcal{M}, \mathcal{R}, \mathcal{D}, \mathcal{U}_{u}, \mathcal{U}_{b}$.}}
\end{algorithmic}
\end{algorithm}

The first step is to verify the number of users per cell interested in the eMBB service. If this number is equal to 1, then the cell will not belong to the MBSFN Area and the user will receive the content through a unicast link. Otherwise, if the number of users exceeds 1, then the considered cell will be part of the set $\mathcal{P}$ of potential cells that could join the future MBSFN Area, wherein users could be served through the MBSFN transmission (lines 1-8). 
{Hence, the considered cell is added to the set $\mathcal{P}$ (line 5) only if the number of users requiring the same content in the given cell is at least equal to 2 since the multicast transmission makes sense if the content has to be delivered to more than 1 user.}

Once the set $\mathcal{P}$ is defined, the algorithm proceeds by determining the set $\mathcal{M}$ that includes all MBSFN Areas, each consisting of adjacent cells (lines 9-11). At this point, the basic MBSFN configuration composed of $\mathcal{M}$, $\mathcal{U}_{b}$, and $\mathcal{U}_{u}$ is generated. Then, the algorithm computes the aggregate data rate of the considered basic configuration (line 12). 

The following step is to sort the MCS supported by $\mathcal{U}_{b}$ in ascending order; after that, starting from the lowest MCS value, the D2D-MAF algorithm begins the evolution process of the basic MBSFN configuration (lines 13-47). In particular, users supporting the $l^{th}$ MCS level are grouped into the set $\mathcal{U}_{l}$ and removed from $\mathcal{U}_{b}$ (lines 15-16). As a consequence, the new $\mathcal{U}_{b}$ is composed of users supporting an MCS level greater than the considered one (i.e., $l$). Then, {to find} the sets $\mathcal{R}$ of relays and $\mathcal{D}$ of D2D receivers among the users in the new $\mathcal{U}_{b}$ and $\mathcal{U}_{l}$, {respectively}, {each gNB of the MBSFN Area follows the standard procedure for the CQI estimation \cite{DCM} and computes its own D2D CSI matrix where CQI feedbacks are reported between potential relays (i.e., the set $\mathcal{U}_b$ is represented by matrix rows) and the users excluded from the MBSFN transmission (i.e., the set $\mathcal{U}_l$ is represented by matrix columns).
If the value of the $b$-th row and the $l$-th column is equal to zero, a D2D connection cannot be activated between the $b$-th UE served by the MBSFN Area and the $l$-th UE outside the MBSFN Area. Otherwise, the $b$-th UE can forward the content received by the gNB to the $l$-th UE. 
According to the values of the D2D CSI matrix, all gNBs select their relays and their subsets of D2D UEs to be associated with each relay. If the $b$-th UE in $\mathcal{U}_b$ has at least one D2D CSI value greater than zero, it is selected as relay and can establish a D2D connection with the $l$-th UE in $\mathcal{U}_l$ (lines 20-21).}

The remaining users of $\mathcal{U}_{l}$ will receive the service via unicast connections (line 22). After having properly allocated resources to unicast users, D2D-MAF algorithm computes the aggregate data rate of the new MBSFN configuration with $\tilde{\mathcal{M}}$, $\tilde{\mathcal{U}}_{b}$, $\tilde{\mathcal{U}}_{u}$ and $\mathcal{D}$ (lines 23-37). If the aggregate data rate of the new MBSFN configuration is greater than the one of the basic MBSFN configuration, then the new MBSFN configuration becomes the new reference MBSFN configuration and the basic one is discarded; otherwise, the algorithm stops (lines 38-46).
After its complete execution, D2D-MAF algorithm provides the best set $\mathcal{M}$ of MBSFN Areas, the configuration of users for each kind of transmission mode (i.e., $\mathcal{U}_{b}$, $\mathcal{U}_{u}$, $\mathcal{D}$ and $\mathcal{R}$) and the related $\mathcal{ADR}$ (line 48). 

\subsection{Complexity of D2D-MAF algorithm}

Below, we provide a detailed analysis of the computational complexity related to the D2D-MAF algorithm.

In lines 1-8, {the algorithm counts how many users are interested in the broadcasted content in order to decide whether a user will be served through either a unicast link or MBSFN transmission by cells belonging to the MBSFN Area. To this aim, the algorithm flows the whole vector of users, hence,} the complexity in finding the number of users interested in the eMBB service is O($|\mathcal{U}|$), where $|\mathcal{U}|$ is the number of users in the system.

In lines 9-11, the creation of MBSFN Areas has a complexity of O($|\mathcal{C}|^2$), where $|\mathcal{C}|$ is the number of cells in the synchronization area. {This is because the operation of MBSFN Area formation mainly depends on the verification of the adjacency constraint among cells. Indeed, the algorithm finds adjacent cells to be included in the same MBSFN Area by analyzing all elements of the adjacency matrix. The adjacency matrix is a $|\mathcal{C}|\times|\mathcal{C}|$ binary matrix, whose elements can assume the value 1 if two cells are adjacent; otherwise, 0.}

In line 12, the ADR computation includes the RB assignment to the MBSFN Area and the D2D connections with a complexity of O($|\mathcal{C}|$). {The ADR computation in itself has constant complexity, but the included radio resource allocation operation is executed as many times as the number of MBSFN Areas. In the worst case, each cell is an MBSFN Area; hence, the number of created MBSFN Areas is at most  $|\mathcal{C}|$.}

In line 13, the sort of the user MCS has a complexity of O($|\mathcal{U}|$ $log(|\mathcal{U}|)$).

In lines 14-47, finding the new MBSFN configuration includes the search for the sets $\mathcal{U}_b$, $\mathcal{R}$, $\mathcal{D}$ and $\mathcal{U}_u$ with the computation of the associated ADR. The overall complexity of these operations is O($|\mathcal{C}|^2\cdot(|\mathcal{U}|^2+|\mathcal{C}|)$). {In particular, the two for-loops (lines 17-18) give the complexity contribution of O($|\mathcal{C}|^2|$). Identifying relays and D2D user has a complexity of O($|\mathcal{U}|^2|$) because the algorithm computes this operation by searching relay and D2D nodes within the D2D CSI matrix that is a $|\mathcal{U}|\times|\mathcal{U}|$ matrix, whose elements can assume a value between 1 and 15 if a D2D connection can be established between two users; otherwise, 0. Finally, the complexity contribution of O($|\mathcal{C}|$) is due to the computation of the new ADR. }
The process to determine the best MBSFN configuration, which improves the system ADR, runs at most \textit{l} times, where \textit{l}=15 is the number of MCS levels.

The implementation of the D2D-MAF algorithm has a polynomial complexity of O($|\mathcal{C}|^2|\mathcal{U}|^2+|\mathcal{C}|^3$), reasonable for real-world scenarios because it is executed by high performing gNBs in a feasible runtime.

\section{Performance Analysis}\label{performance evaluation}

\subsection{Simulative model}

To assess the effectiveness of the proposed D2D-MAF algorithm, simulations have been performed by {an ad-hoc developed simulator in MATLAB, specifically designed for the MBSFN Area formation.} Each simulation has been run several times to achieve the most reliable results with 95\% confidence intervals.

We consider a synchronization area including a number of cells ranging from 10 to 36, each one with a coverage radius of 250 m. 
Users are randomly deployed within each cell of the synchronization area. Moreover, we consider that cells of the same MBSFN Area constructively interfere, otherwise they are sources of disruptive interference. The main simulation settings are reported in Table \ref{tab:parameter}.

\begin{table}[ht]
\caption{Main Simulation Assumptions}\label{tab:parameter}
\centering{
\begin{tabular}{|c|c|}\hline
\textbf{Parameter} & \textbf{Value} \\\hline
Cell layout & Hexagonal grid, 10 cells\\\hline
Inter Site Distance & 500 m\\\hline 
Pathloss model & 128.1+37.6 $log_{10}$(R),
R in kilometers \cite{systemmodel_standard}\\\hline
{gNB} transmit power & 46 dBm\\\hline
D2D node Tx power&  23 dBm\\\hline
{gNB} antenna gain & 15 dBi\\\hline
UE antenna gain & 0 dBi\\\hline
{gNB} noise figure & 5 dB\\\hline
UE noise figure & 9 dB\\\hline
Carrier frequency & 2 GHz\\\hline
Scheduling Frame & 10 ms\\\hline
RB size & 12 sub-carrier\\\hline
{$\mu$} &  {0}\\\hline
Sub-carrier spacing & 15 kHz\\\hline
TTI & 1 ms\\\hline
BLER target & 1\% \\\hline
\end{tabular}}
\end{table}

The performance of the D2D-MAF algorithm is compared to that of the SCF algorithm \cite{8} \cite{SCF_2017}.  
For a fair comparison between D2D-MAF and SCF, we make the following two assumptions.
First, the two algorithms adopt the same resource allocation strategy. 
In particular, we assume that both algorithms serve users, interested in the eMBB service, via unicast links by exploiting Inband resources \cite{BMSB2018}. This means that these unicast transmissions take their radio resources from the pool of RBs destined to the MBSFN transmission.
Second, both D2D-MAF and SCF algorithms always respect the constraint of satisfying 100\% of user requests.

We focus our attention on three simulation scenarios:
\begin{itemize}
\item \textit{Scenario 1}, where the number of users per cell varies from 200 to 400 and the channel bandwidth is fixed to {50} MHz (i.e., {270} RBs) {\cite{NR_BW}}.
\item \textit{Scenario 2}, where the number of users per cell is set to 300, the channel bandwidth is {50} MHz, and the number of cells varies from 10 to 36.
\item \textit{Scenario 3}, where the number of users per cell is fixed to 300 and the channel bandwidth varies from {5} to {50} MHz.
\end{itemize}

In all simulation scenarios, we analyze the performance of D2D-MAF and SCF algorithms also under different TDD frame configurations (see Table \ref{tab:frames}).

The performance metrics under investigation are:
\begin{itemize}
\item \textit{Average Throughput} is the average data rate experienced by all users.
\item \textit{Aggregate Data Rate} (ADR) is computed as the sum of each user throughput.
\item \textit{Delivery Time} is the average time needed by all users to receive a content of 20 MB.
\item \textit{Used RB for D2D} is the percentage of radio resources exploited for the D2D transmission in the uplink subframe.
\end{itemize}

\subsection{Results analysis}

\subsubsection{Scenario 1}
\begin{figure*}[ht]
\centering
\subfigure[{TDD conf. 0}]{\includegraphics[scale=0.3]{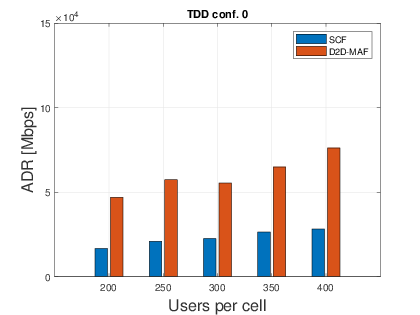}}
\label{fig:ADR_UE_frame0}
\subfigure[{TDD conf. 1}]{\includegraphics[scale=0.3]{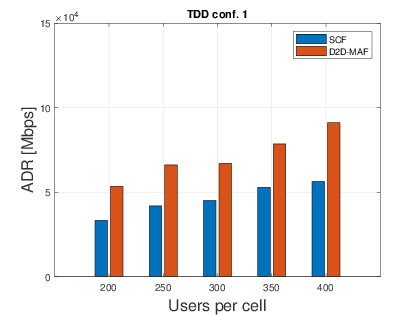}}
\label{fig:ADR_UE_frame1}
\subfigure[{TDD conf. 2}]{\includegraphics[scale=0.3]{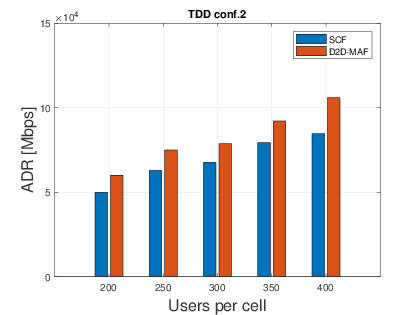}}
\label{fig:ADR_UE_frame2}
\subfigure[{TDD conf. 3}]{\includegraphics[scale=0.3]{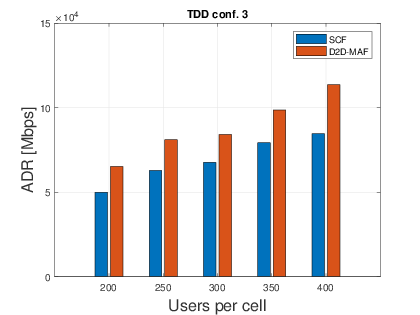}}
\label{fig:ADR_UE_frame3}
\subfigure[{TDD conf. 4}]{\includegraphics[scale=0.3]{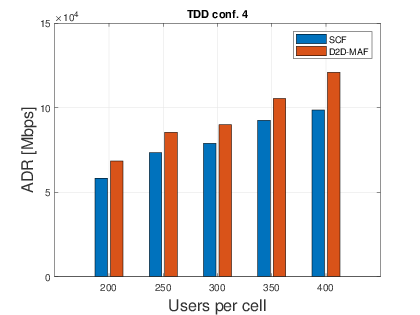}}
\label{fig:ADR_UE_frame4}
\subfigure[{TDD conf. 5}]{\includegraphics[scale=0.3]{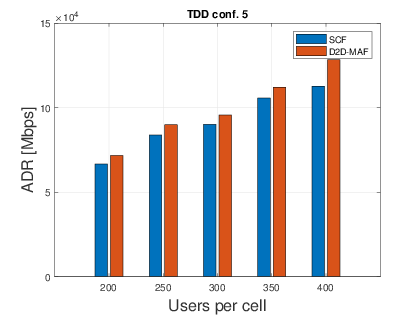}}
\label{fig:ADR_UE_frame5}
\subfigure[{TDD conf. 6}]{\includegraphics[scale=0.3]{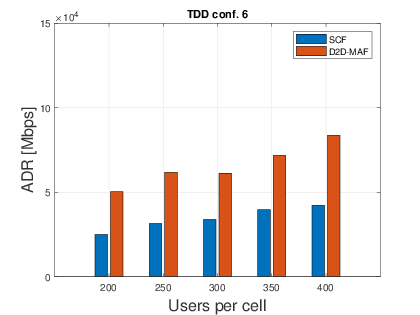}}
\label{fig:ADR_UE_frame6}
\caption{{ADR under varying number of users per cell for (a) TDD conf. 0, (b) TDD conf. 1, (c) TDD conf. 2, (d) TDD conf. 3, (e) TDD conf. 4, (f) TDD conf. 5, (g) TDD conf. 6.}}
\label{fig:ADR_UE}
\end{figure*}

\begin{figure*}[ht!]
\centering
\subfigure[{D2D-MAF}]{\includegraphics[scale=0.48]{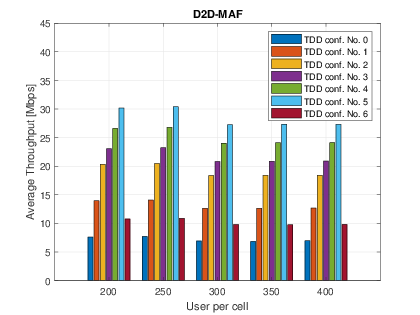}}
\label{fig:THR_UE_D2DMAF}
\subfigure[{SCF}]{\includegraphics[scale=0.48]{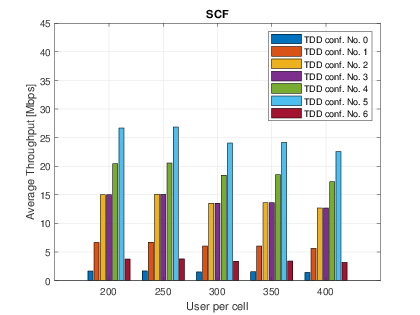}}
\label{fig:THR_UE_SCF}
\caption{{Average Throughput of users under varying number of users per cell for (a) D2D-MAF, and (b) SCF.}}
\label{fig:THR_UE}
\end{figure*}
\begin{figure*}[ht!]
\centering
\subfigure[{D2D-MAF}]{\includegraphics[scale=0.48]{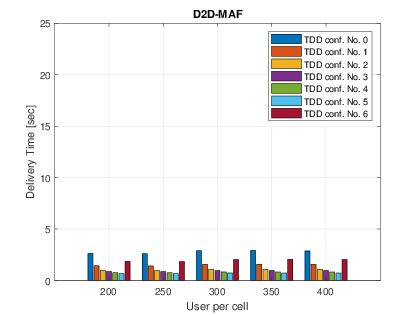}}
\label{fig:DT_UE_D2DMAF}
\subfigure[{SCF}]{\includegraphics[scale=0.48]{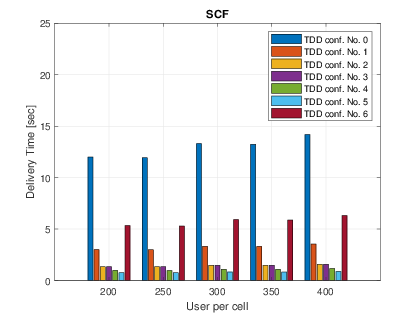}}
\label{fig:DT_UE_SCF}
\caption{{Delivery Time under varying number of users per cell for (a) D2D-MAF, and (b) SCF.}}
\label{fig:DT_UE}
\end{figure*}


In this scenario, we assess how D2D-MAF and SCF algorithms behave when varying the number of users per cell (varying from 200 to 400). We analyze the system performance in terms of ADR (Fig. {\ref{fig:ADR_UE}}), average throughput (Fig. {\ref{fig:THR_UE}}) and delivery time (Fig. {\ref{fig:DT_UE}}) for seven TDD frame configurations.
As expected, the proposed D2D-MAF shows better performance than SCF for all the considered TDD frame configurations.
Generally, the ADR  is improved when the number of users per cell increases. This is due to the cumulative nature of this metric. However, the improvement is ``slight'' because the higher the number of users, the higher the probability to find users with poorest channel conditions. Due to this motivation, the average throughput lightly decreases when the number of users in the system gets greater. As a consequence, the delivery time gets a little longer.
However, performing the proposed D2D-MAF allows the whole service to be delivered in a shorter time compared to SCF. 

We highlight that the best results are achieved when considering the TDD frame configuration No. 5. Indeed, the large number of downlink subframes allows buffering a high amount of data to be forwarded during the uplink subframes. In this case, the potential of the D2D link is exploited to the maximum by providing a lower delivery time and a higher throughput, which means, as a consequence, a higher ADR with respect to other TDD frame configurations.
In fact, thanks to the good channel quality perceived on D2D links due to proximity of communicating devices, the transmission between D2D relays and D2D receivers can take advantage from less robust MCSs with respect to the more robust MCS selection required by the gNB-to-D2D relay transmission (i.e., the time required to transmit the same amount of data to D2D relays is substantially higher than the one needed on D2D links). Thus, the whole UL subframe can be exploited for relaying stored data received from the gNB. In TDD configurations with a higher number of UL subframes, it is likely that some subframes are either only partially exploited (not all available RBs are occupied) or not utilized at all in case not enough data has been delivered from gNB to D2D relay nodes.
In fact, the TDD frame configuration No. 0 shows the worst performance, due to the presence of more uplink than downlink subframes.

\subsubsection{Scenario 2}

In this scenario, we analyze D2D-MAF and SCF algorithm behaviors for all the considered TDD frame configuration when increasing the size of the MBSFN Area (i.e., the number of cells).
As expected, the proposed D2D-MAF algorithm achieves a higher ADR and an average throughput than the SCF scheme, even in this scenario. 

\begin{figure*}[h]
\centering
\subfigure[{TDD conf. 0}]{\includegraphics[scale=0.30]{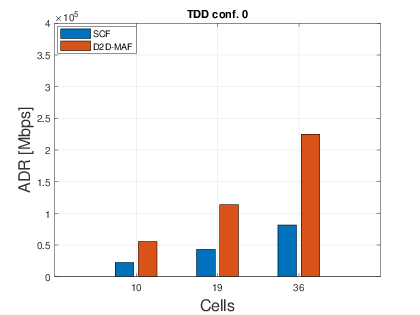}}
\label{fig:ADR_CELLE_NR_frame0}
\subfigure[{TDD conf. 1}]{\includegraphics[scale=0.30]{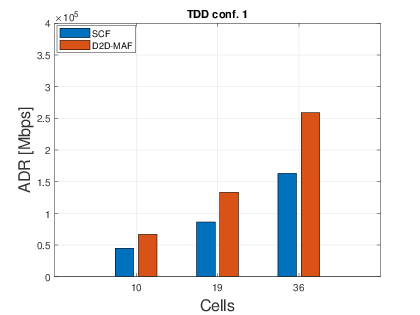}}
\label{fig:ADR_CELLE_NR_frame1}
\subfigure[{TDD conf. 2}]{\includegraphics[scale=0.30]{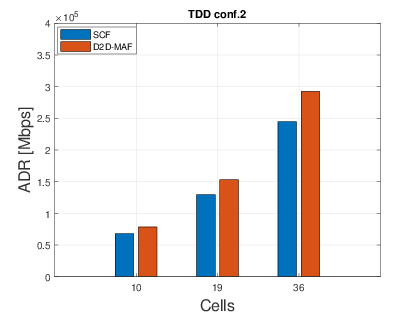}}
\label{fig:ADR_CELLE_NR_frame2}
\subfigure[{TDD conf. 3}]{\includegraphics[scale=0.30]{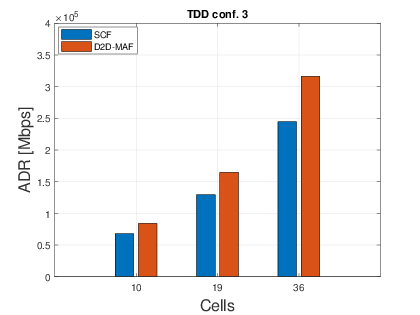}}
\label{fig:ADR_CELLE_NR_frame3}
\subfigure[{TDD conf. 4}]{\includegraphics[scale=0.30]{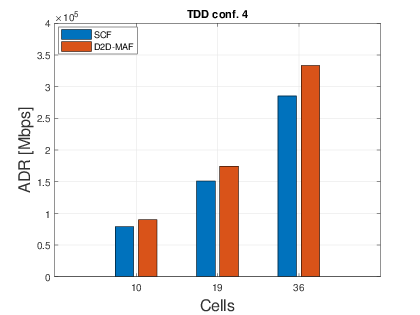}}
\label{fig:ADR_CELLE_NR_frame4}
\subfigure[{TDD conf. 5}]{\includegraphics[scale=0.30]{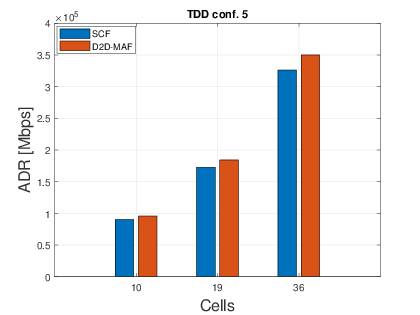}}
\label{fig:ADR_CELLE_NR_frame5}
\subfigure[{TDD conf. 6}]{\includegraphics[scale=0.30]{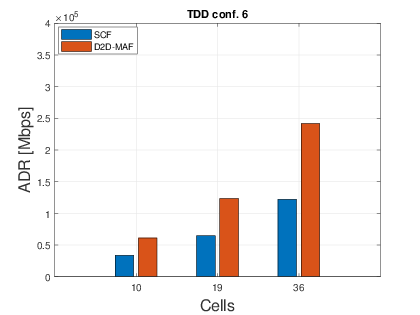}}
\label{fig:ADR_CELLE_NR_frame6}
\caption{{ADR under varying number of cells within the synchronization area for (a) TDD conf. 0, (b) TDD conf. 1, (c) TDD conf. 2, (d) TDD conf. 3, (e) TDD conf. 4, (f) TDD conf. 5, (g) TDD conf. 6.}}
\label{fig:ADR_CELLE}
\end{figure*}

In detail, by looking at Fig. \ref{fig:ADR_CELLE} we note that the ADR increases when considering a fixed number of users per cell and an ever-larger number of cells within the MBSFN Area. This is due to the cumulative nature of this metric and, hence, to the contribution of each user. 
\begin{figure*}[h!]
\centering
\subfigure[{D2D-MAF}]{\includegraphics[scale=0.48]{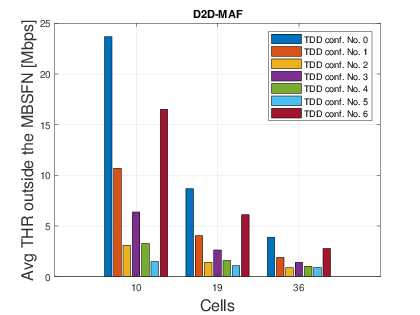}}
\label{fig:THR_CELLE_D2DMAF}\quad
\subfigure[{SCF}]{\includegraphics[scale=0.48]{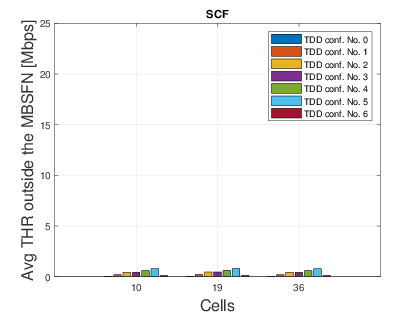}}
\label{fig:THR_CELLE_SCF}
\caption{{Average Throughput of users outside the MBSFN Area under varying number of cells within the synchronization area for (a) D2D-MAF, and (b) SCF.}}
\label{fig:THR_CELLE}
\end{figure*}
Fig. \ref{fig:THR_CELLE} shows the average throughput delivered to those users outside the MBSFN Area that receive the content through either D2D or unicast links. In D2D-MAF (Fig. \ref{fig:THR_CELLE}(a)), cell-edge users take advantage from the D2D connectivity in single frequency mode with respect to the traditional unicast connection exploited in SCF (Fig. \ref{fig:THR_CELLE}(b)). In fact, their data rates significantly increase thanks to the possibility to establish high performing D2D links between cell-edge users and forwarding devices perceiving better channel conditions than those experienced towards the gNB.
However, the average throughput decreases when the number of cells grows. Indeed, the overall number of users in the system increases with the number of cells since we consider a fixed number of users per cell. Consequently, the probability of having users with the poorest channel conditions that influence the choice of the multicast transmission parameters (i.e., more robust modulation) is higher.
Nevertheless, the system performance in terms of ADR and average throughput results to be better for the proposed D2D-MAF than for the SCF algorithm.

Furthermore, both the algorithms experience the worst and the best performance conditions, in terms of ADR and average throughput, with the TDD frame configuration No. 0 and No. 5, respectively. We remark that, while for SCF downlink subframes are the only opportunity to transmit the content, D2D-MAF exploits also uplink subframes to allocate resources to D2D transmissions and improve the system performance. {Therefore, in TDD frame configuration with a few D subframes (i.e., TDD frame configuration No. 0) most of the U subframes are not exploited because of no content to forward (i.e., relay device buffer is empty). On the contrary, in TDD frame configurations with many D subframes, U subframes are efficiently used to relay all data accumulated in the relay device buffer. This is taken to the extremes is TDD frame configuration No. 5, wherein only one U subframe can be exploited for D2D data delivery. To speed up the buffer emptying and to deliver content to D2D users as quickly as possible, D2D-MAF foresees that all available D2D radio resources are allocated for the D2D transmission in that U subframe, thus leading to the full exploitation of D2D link potentialities (i.e., very high-performing link thanks to extremely high data rates).}

{The ADR is computed as the sum of all users’ data rates on a frame-basis, hence, the ADR contribution of users receiving the content through MBSFN and unicast transmissions is given in D subframes; whereas the ADR contribution of users receiving the content via D2D links is given in U subframes. Due to the presence of only one U subframe against the eight D subframes, in TDD frame configuration No. 5 the data rate contribution of D2D users is lower than the D2D ADR contribution in TDD frame configuration No. 1 (where there is the same number of U and D subframes within a frame). This is why the performance gap between the proposed D2D-MAF and SCF becomes smaller in a TDD frame configuration with a number of U subframes lower than the number of D subframes. Despite the small gap (i.e., about 7\% on average in Fig. \ref{fig:ADR_CELLE}(f)) the TDD frame configuration No. 5 provides the best ADR performance results (i.e., higher ADR with respect to other TDD frame configurations) in all considered scenarios.}


\subsubsection{Scenario 3}

{In this scenario, we evaluate the behavior of both D2D-MAF and SCF algorithms under 5G NR frequency bands}. 
{The 5G NR technology extends the LTE spectrum by introducing new frequencies below 1 GHz to 52.6 GHz. In particular, the frequency range FR1 includes all existing and new bands below 7 GHz while FR2 includes new bands in the range from 24.25 GHz to 52.6 GHz \cite{NR_BW}.}
{For all the considered TDD frame configurations, we analyze the system performance when considering numerology $\mu=0$ and subcarrier spacing $SCS=15$ kHz supported only by the FR1 \cite{NR_BW}. NR channel bandwidths specified for numerology $\mu=0$ are 5, 10, 15, 20, 25, 30, 40, and 50 MHz.}

\begin{figure*}[ht!]
\centering
\subfigure[{TDD conf. 0}]{\includegraphics[scale=0.3]{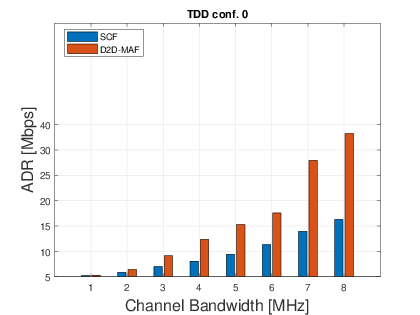}}
\label{fig:ADR_BW_frame0}
\subfigure[{TDD conf. 1}]{\includegraphics[scale=0.3]{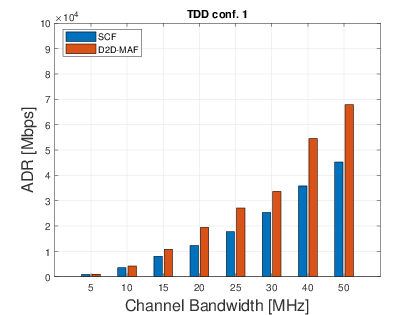}}
\label{fig:ADR_BW_frame1}
\subfigure[{TDD conf. 2}]{\includegraphics[scale=0.3]{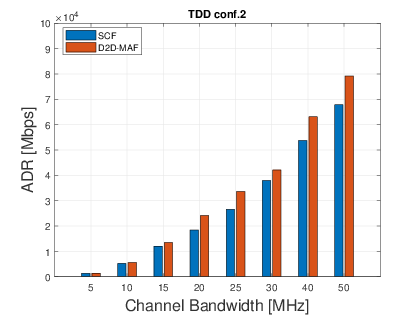}}
\label{fig:ADR_BW_frame2}
\subfigure[{TDD conf. 3}]{\includegraphics[scale=0.3]{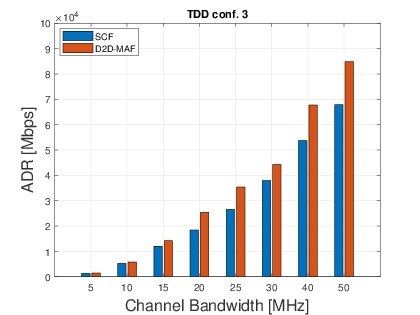}}
\label{fig:ADR_BW_frame3}
\subfigure[{TDD conf. 4}]{\includegraphics[scale=0.3]{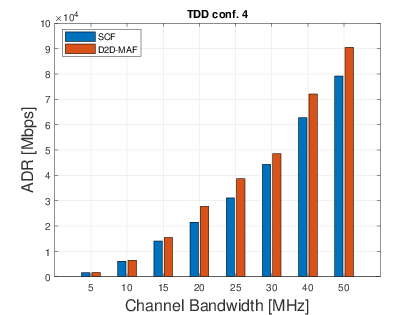}}
\label{fig:ADR_BW_frame4}
\subfigure[{TDD conf. 5}]{\includegraphics[scale=0.3]{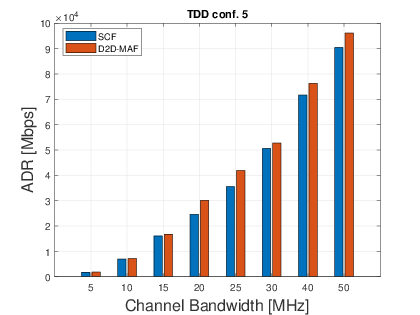}}
\label{fig:ADR_BW_frame5}
\subfigure[{TDD conf. 6}]{\includegraphics[scale=0.3]{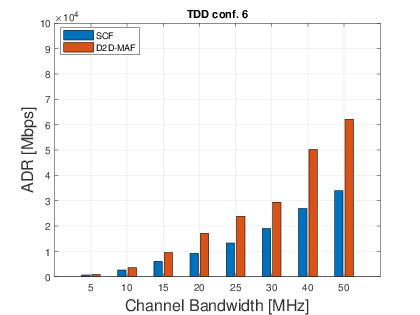}}
\label{fig:ADR_BW_frame6}
\caption{{ADR under varying NR channel bandwidth given (a) TDD conf. 0, (b) TDD conf. 1, (c) TDD conf. 2, (d) TDD conf. 3, (e) TDD conf. 4, (f) TDD conf. 5, (g) TDD conf. 6.}}
\label{fig:ADR_RB}
\end{figure*}

\begin{figure*}[ht!]
\centering
\subfigure[{D2D-MAF}]{\includegraphics[scale=0.48]{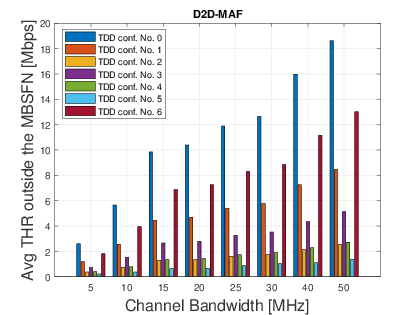}}
\label{fig:THR_BW_D2DMAF}\quad
\subfigure[{SCF}]{\includegraphics[scale=0.48]{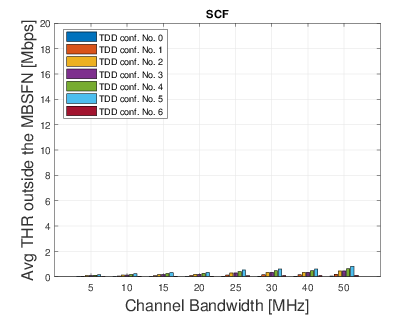}}
\label{fig:THR_BW_SCF}
\caption{{Average Throughput of users outside the MBSFN Area under varying NR channel bandwidth for (a) D2D-MAF and (b) SCF.}}
\label{fig:THR_RB}
\end{figure*}

\begin{figure*}[ht!]
\centering
\subfigure[{D2D-MAF}]{\includegraphics[scale=0.48]{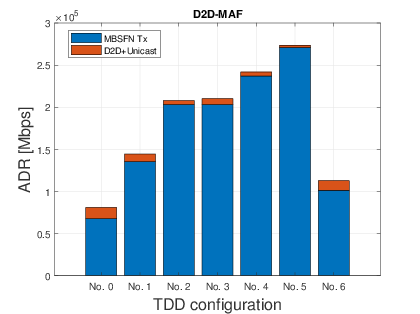}}
\label{fig:ADR_FRAME_D2DMAF}
\subfigure[{SCF}]{\includegraphics[scale=0.48]{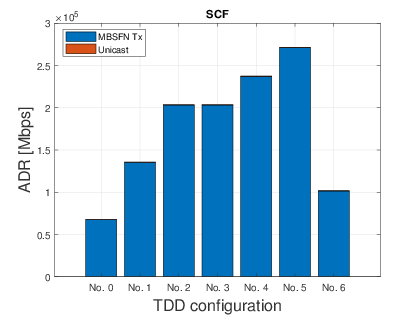}}
\label{fig:ADR_FRAME_SCF}
\caption{{ADR contribution due to the MBSFN Transmission and (a) both unicast and D2D links for D2D-MAF and (b) only unicast for SCF.}}
\label{fig:FRAME}
\end{figure*}

As expected, under increasing channel bandwidth, both ADR (Fig. \ref{fig:ADR_RB}) and average throughput of users outside the MBSFN Area (Fig. \ref{fig:THR_RB}) get higher for both D2D-MAF and SCF algorithms.
The reason for this behavior lies in a selection of transmission parameters (i.e., modulation and coding scheme) which depends on the number of available radio resources. In detail, when a few RBs are available in the system, only a limited number of users can be served via unicast links. This means that the two algorithms select a more robust, thus less performing, MCS level with respect to the case when a wider channel bandwidth (i.e., more RBs) is available. However, D2D-MAF always outperforms SCF because the exploitation of D2D communications makes the difference in terms of system performance. Indeed, a considerable portion of cell-edge users perceives an improved quality of service because of the high-performing D2D links established with forwarding devices under the coverage of the MBSFN Area. Hence, users with poor channel conditions improve their experience without affecting the MBSFN transmission towards devices perceiving a good channel quality. Unicast connections are established only as a last chance to deliver multicast content.
In this regard, Fig. \ref{fig:FRAME} shows the contribution of ADR for D2D-MAF (see Fig. \ref{fig:FRAME}(a)) and SCF (see Fig. \ref{fig:FRAME}(b)) due to MBSFN and non-multicast (both unicast and D2D for D2D-MAF, only unicast for SCF) connections. The introduction of D2D communications that support the eMBB service delivery to users outside the MBSFN Area leads to two important benefits. First, the coverage radius of the MBSFN Area is reduced with a consequent choice of a less robust modulation during transmission. This implies the improvement of the perceived quality of service and the increase of user experienced data rate and system ADR. Second, the contribution in terms of ADR given by the usage of D2D links is up to 100 times higher than the case when only unicast transmissions are established.

\begin{figure}[ht!]
\centering
{\includegraphics[scale=0.45]{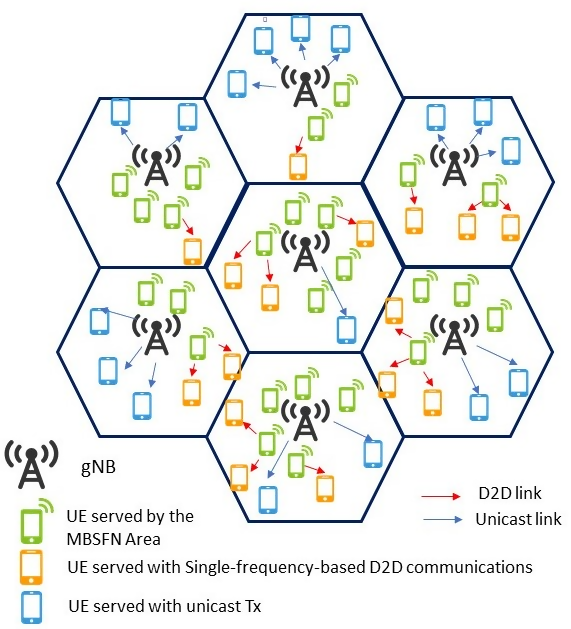}}
\caption{{Percentage of used RBs for D2D in all the considered TDD configurations when varying NR channel bandwidth.}}
\label{fig:usedRBD2D_RB}
\end{figure}

As shown above (from Fig. 2 to Fig. 8), the best performance levels (highest ADR and throughput, and lowest delivery time) is achieved under TDD frame configuration No. 5, which has the lowest number of uplink subframes. On the opposite, TDD frame configuration No. 0, gathering the highest number of uplink subframes, exhibits the worst performance levels.
In case of a low number of uplink subframes in a frame compared to the number of downlink subframes, the potential of D2D is fully exploited. In fact, available uplink radio resources are assigned to D2D transmission in order to allow forwarding devices to deliver to cell-edge users the data received by gNB in previous downlink subframes. When considering the TDD frame configuration No. 5, a higher number of RBs are exploited by D2D compared to those used in TDD frame configuration No. 0. In addition, when considering TDD frame configurations with an equal number of uplink subframes (TDD conf. No. 4 and No. 2), we can appreciate a difference in terms of percentage of used RB for D2D due to the position of uplink subframes inside the frame. Indeed, when the downlink subframes are consecutive, more uplink radio resources are utilized because a greater amount of data, accumulated in the buffer during downlink subframes, needs to be forwarded to cell-edge users. If downlink subframes are not consecutive in the frame, the buffer of the forwarding devices is partially emptied from time to time, thus requiring fewer RBs with respect to the previous case (i.e. consecutive downlink subframes).
Fig. \ref{fig:usedRBD2D_RB} confirms that both the number of uplink/downlink subframes and their position within the frame are important for determining how many RBs are needed to forward the eMBB content through D2D links.

\section{Conclusions}
\label{conclusion}

In this paper, we propose a D2D-aided MBSFN Area Formation (D2D-MAF) algorithm designed to dynamically select the best configuration of the MBSFN Area, which increases the system Aggregate Data Rate while satisfying all user requests in 5G NR systems. The proposed algorithm reduces the coverage radius of the MBSFN Area by choosing, step by step, a less robust modulation supported only by those users experiencing good channel quality. To serve users with poor channel conditions, D2D-MAF exploits either Device-to-Device (D2D) or unicast transmission by preferring D2D connection whenever a direct link among devices can be established. Simulation campaigns have been conducted to assess the effectiveness of the proposed D2D-MAF algorithm for seven TDD frame configurations in three different scenarios.
Simulative results show that D2D-MAF provides better system performance in terms of ADR, average throughput and delivery time for all the considered TDD frame configurations w.r.t. other compared solutions.
\ifCLASSOPTIONcaptionsoff
  \newpage
\fi

\begin{IEEEbiography}[{\includegraphics[width=1in,height=1.25in,clip,keepaspectratio]{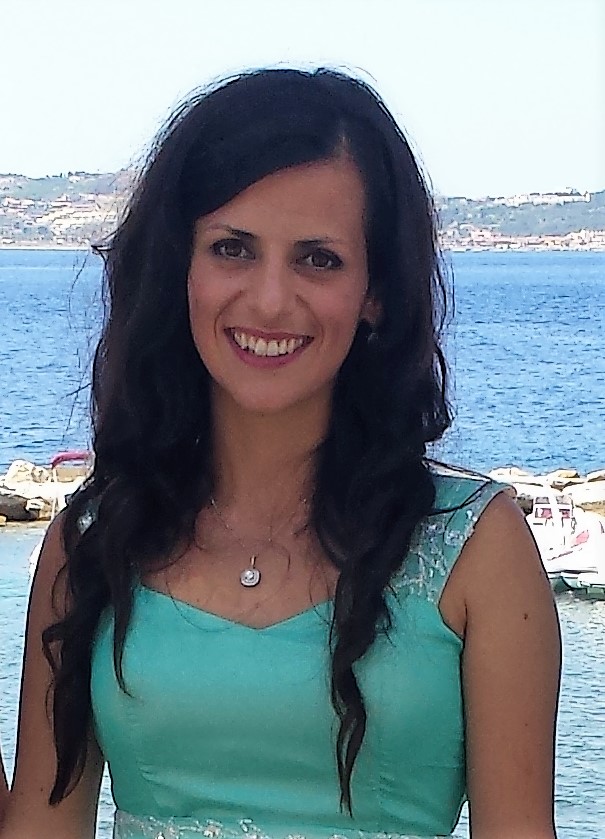}}]{Federica Rinaldi} received the B.Sc. degree in telecommunication engineering and M.Sc. degree \textit{(cum laude)} in computer science and telecommunication systems engineering from the University Mediterranea of Reggio Calabria, Italy, in 2013 and 2017, respectively. She is currently pursuing the Ph.D. degree in information engineering. Her current research interests include non-terrestrial networks, radio resource management, multimedia broadcast and multicast service, and device-to-device communications over 5G networks.
\end{IEEEbiography}

\begin{IEEEbiography}[{\includegraphics[width=1in,height=1.25in,clip,keepaspectratio]{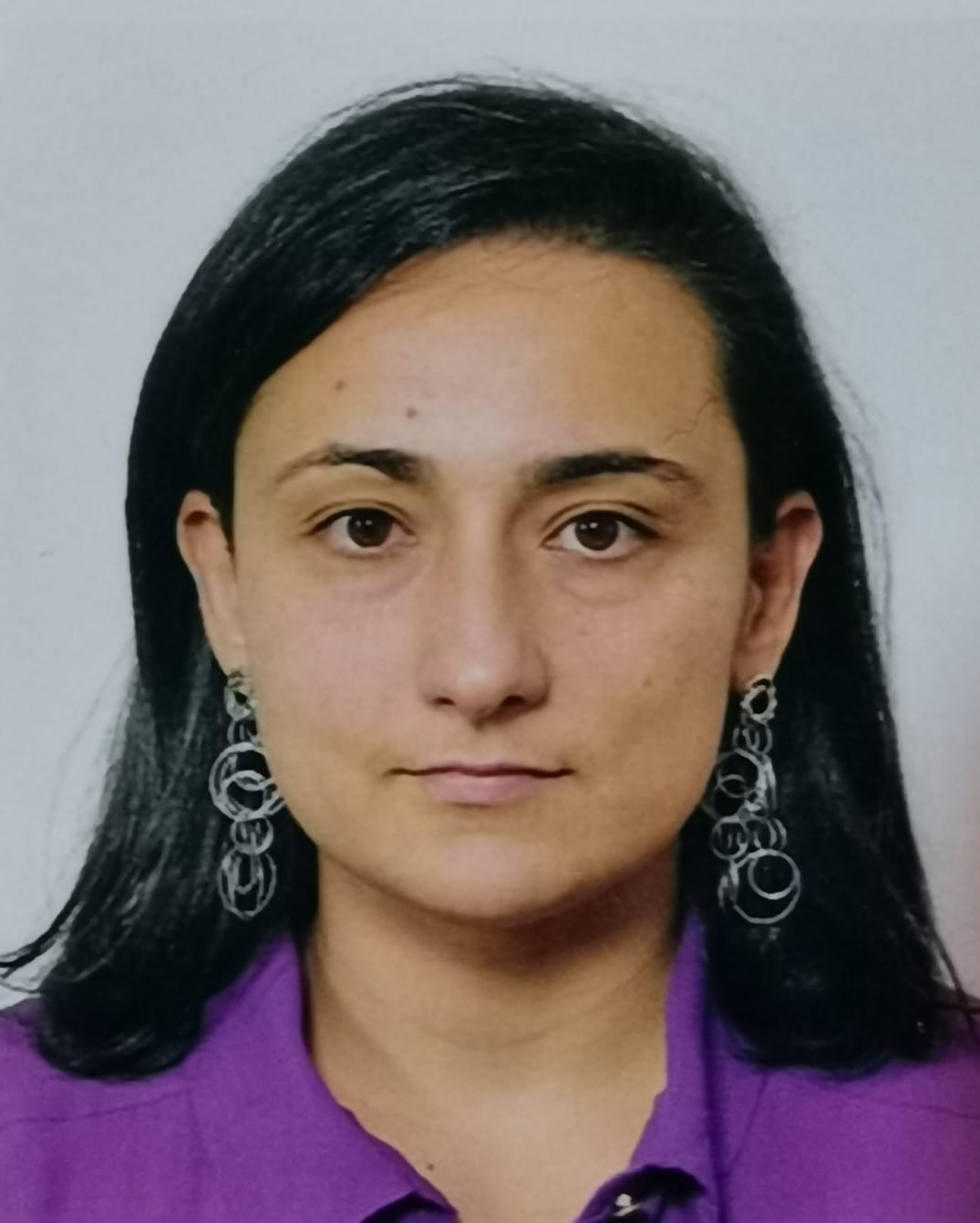}}]{Sara Pizzi} is Research Assistant in Telecommunications at University Mediterranea of Reggio Calabria, Italy. From the same University she received the 1st (2002) and 2nd (2005) level Laurea Degree, both cum laude, in Telecommunication Engineering and the Ph.D. degree (2009) in Computer, Biomedical and Telecommunication Engineering. In 2005, she received a Master's degree in Information Technology from CEFRIEL/Politecnico di Milano. Her current research interests focus on radio resource management for multicast service delivery, Device-to-Device and Machine Type Communications over 5G networks, integration of Non-Terrestrial Networks in Internet of Things.
\end{IEEEbiography}

\begin{IEEEbiography}[{\includegraphics[width=1in,height=1.25in,clip,keepaspectratio]{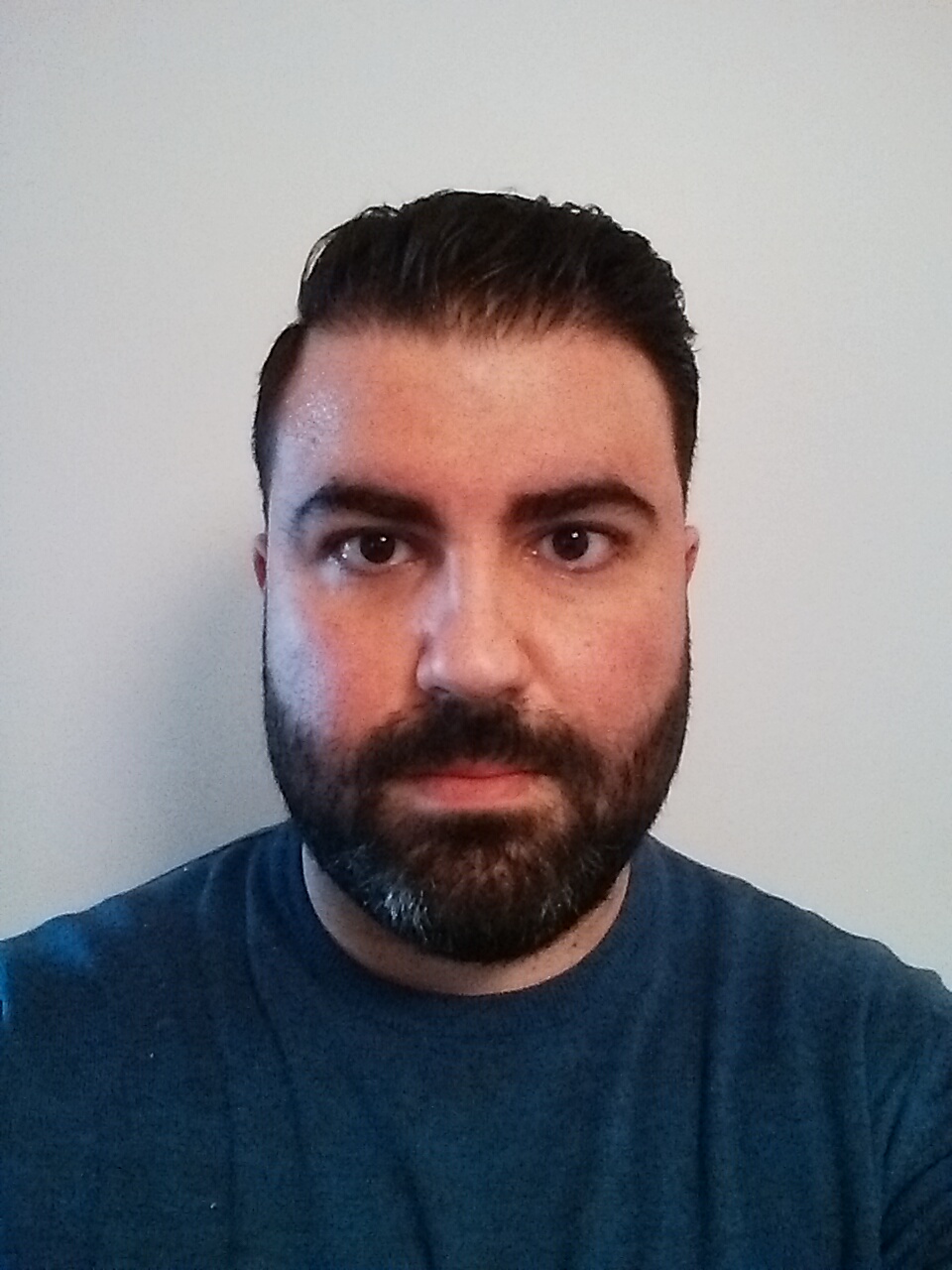}}]{Antonino Orsino} is currently a Senior Researcher at Ericsson Research, Finland, and an Ericsson 3GPP delegate in the RAN2 WG. He received the B.Sc. degrees in Telecommunications Engineering from University Mediterranean of Reggio Calabria, Italy, in 2009 and the M.Sc. from University of Padova, Italy, in 2012. He also received his Ph.D. from University Mediterranea of Reggio Calabria, Italy, in 2017. He is actively working in 5G NR standardization activities and additional research interests include Device-to-Device and Machine-to-Machine communications in 4G/5G cellular systems, and Internet of Things. He is the inventor/co-inventor of 50+ patent families, as well as the author/co-author of 60+ international scientific publications and numerous standardization contributions in the field of wireless networks. He received the Best Junior Carassa Award in 2016 as the best Italian junior researcher in Telecommunications. He has been a co-organizer of the GET-IoT workshop co-located with the European Wireless 2017 and 2018 conference and co-chair of the Wireless Networking and Multimedia symposium within the IEEE/CIC ICCC 2018. He is currently an Associate Editor for the IEEE Access journal and has served as TPC member and designated reviewer in many international IEEE conferences and journals.
\end{IEEEbiography}

\begin{IEEEbiography}[{\includegraphics[width=1in,height=1.20in,clip,keepaspectratio]{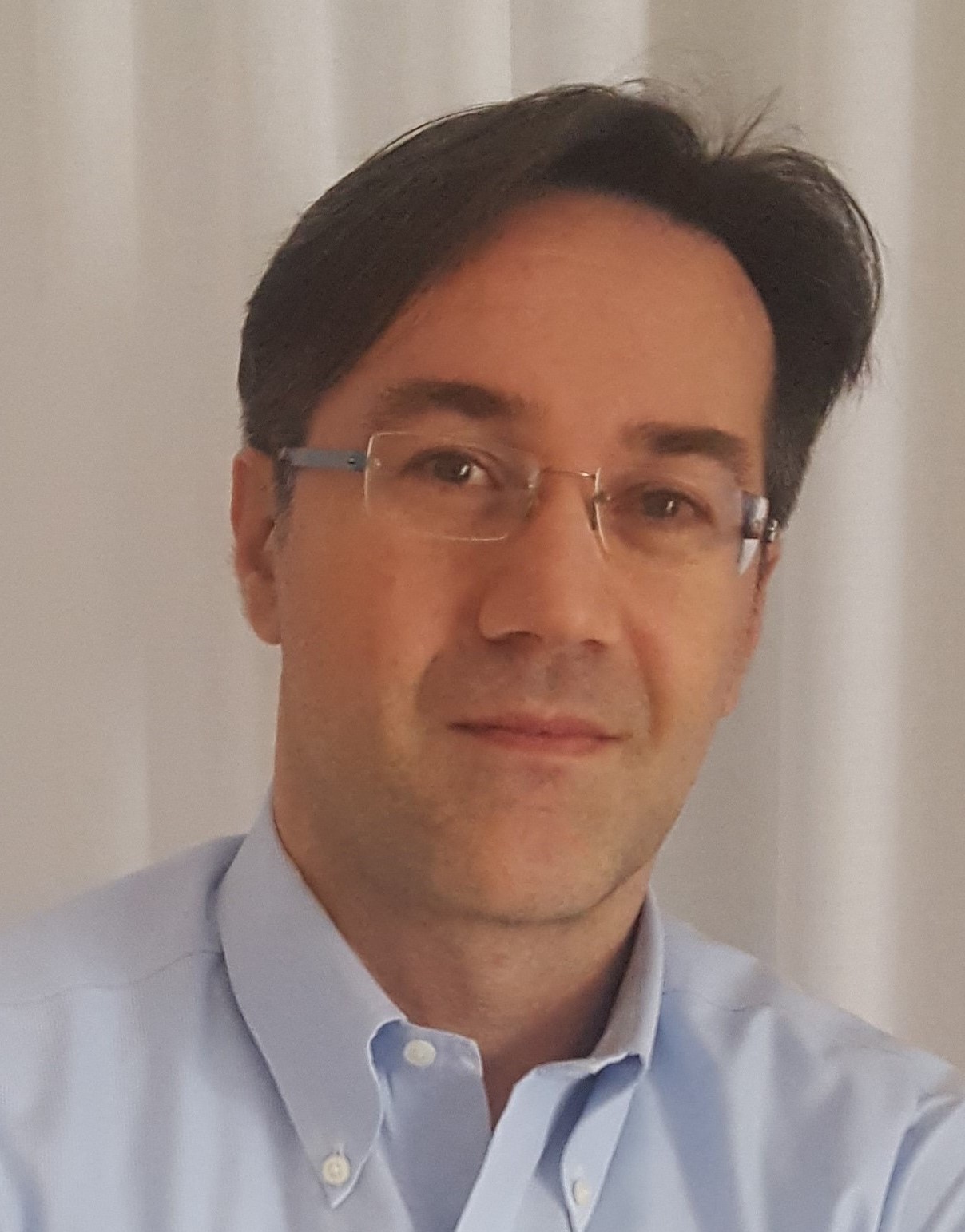}}]{Antonio Iera} graduated in computer engineering from the University of Calabria in 1991, and received a Master's degree in IT from CEFRIEL/Politecnico di Milano in 1992 and a Ph.D. degree from the University of Calabria in 1996. From 1997 to 2019 he has been with the University Mediterranea, Italy, and currently holds the position of full professor of Telecommunications at the University of Calabria, Italy. His research interests include next generation mobile and wireless systems, and the Internet of Things.
\end{IEEEbiography}

\begin{IEEEbiography}[{\includegraphics[width=1in,height=1.25in,clip,keepaspectratio]{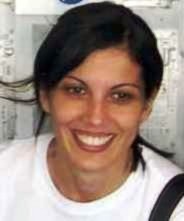}}]{Antonella Molinaro}
graduated in Computer Engineering (1991) at the University of Calabria, received a Master degree in Information Technology from CEFRIEL/Polytechnic of Milano (1992), and a Ph.D. degree in Multimedia Technologies and Communications Systems (1996). She was with Telesoft S.p.A., Rome (1992–1993) and Siemens A.G., Munich, Germany (1994–1995) as a CEC Fellow in the RACE-II program. She was a research fellow at the Polytechnic of Milano (1997–1998), and an assistant professor with the University of Messina (1998–2001) and the University of Calabria (2001–2004). She is currently an associate professor of telecommunications at the University Mediterranea of Reggio Calabria, Italy, and a professor at CentraleSup\'elec--CNRS--Universit\'e Paris-Saclay, France. Her research activity mainly focuses on wireless and mobile networking, vehicular networks, and future Internet.
\end{IEEEbiography}

\begin{IEEEbiography}[{\includegraphics[width=1in,height=1.25in,clip,keepaspectratio]{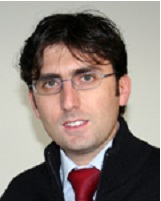}}]{Giuseppe Araniti}
(S'03-M'05-SM'15) received the
Laurea degree and the Ph.D. degree in electronic
engineering from the University Mediterranea of
Reggio Calabria, Italy, in 2000 and 2004, respectively, where he is an Assistant Professor of telecommunications. His major area of research is on 5G networks and it includes personal communications, enhanced wireless and satellite systems, traffic and radio resource management, multicast and broadcast 
services, device-to-device (D2D) and machine-type communications (M2M/MTC). He is an Associate 
Editor of the IEEE Transactions on Broadcasting and of IEEE Transaction on Vehicular Technology, is Vice-Chair of the IEEE BTS Italian 
Chapter and of the Special Interest Group on Social Behaviour Driven Cognitive Radio Networks – IEEE 
Comsoc.
\end{IEEEbiography}
%








\end{document}